\documentclass{PoS}

\title{APEX observations of non-stationary magneto-hydrodynamical shocks in W44}

\ShortTitle{Non-stationary magneto-hydrodynamical shocks in W44}

\author{\speaker{Sibylle Anderl}\\
        Univ. Grenoble Alpes, IPAG, F-38000 Grenoble, France\\
CNRS, IPAG, F-38000 Grenoble, France \\
        E-mail: \email{sibylle.anderl@obs.ujf-grenoble.fr}}

\author{Antoine Gusdorf\\
	LERMA, UMR 8112 du CNRS, Observatoire de Paris, École Normale Supérieure, 24 rue Lhomond, F75231 Paris Cedex 05, France\\
   E-mail: \email{antoine.gusdorf@lra.ens.fr}
    }
      
\author{Rolf G\"usten\\
	Max Planck Institut f\"ur Radioastronomie, 
              Auf dem H\"ugel 69, 53121 Bonn, Germany\\
            E-mail: \email{guesten@mpifr-bonn.mpg.de}
 }

\abstract{Supernova remnants (SNRs) are considered as being the sources of galactic cosmic rays. In order to understand the origin, acceleration, and composition of these cosmic rays, detailed knowledge of the physical conditions in the local interstellar medium is needed. The shock interaction of SNRs with molecular clouds that gives rise to strong molecular emission in the far-IR and sub-mm wavelength regimes can be used as a highly valuable tracer of these conditions. The application of MHD shock models in the interpretation of the resulting line emission can yield information on the energetic and chemical impact of supernova remnants. We have mapped two regions in the supernova remnant W44 with the APEX telescope in $^{12}$CO (3--2), (4--3), (6--5), (7--6) and $^{13}$CO (3--2). The extraction of integrated intensities on five different positions, corresponding to local maxima of CO emission, allows to compare these intensities to the outputs of a grid of models, which combine an MHD shock code with
a radiative transfer module based on the 'large velocity gradient' approximation. We find that the observed CO line emission is compatible with non-stationary shocks and a pre-shock density of 10$^4$ cm$^{-3}$. Our models furthermore allow to constrain shock ages, velocities, the pre-shock magnetic field strength components perpendicular to the line-of-sight, and the full ladder of CO transitions. Finally, our analysis can be used to estimate the contribution of such SNRs to, e.g. the galactic energy balance and the momentum-injection into the surrounding interstellar medium.}

\FullConference{Cosmic Rays and the InterStellar Medium - CRISM 2014,\\
		24-27 June 2014\\
		Montpellier, France}

\begin{document}
\section{Introduction}
The dynamical state of the interstellar medium (ISM) is strongly affected by supernova explosions. Regions of hot and tenuous gas are created when the explosion injects about 10$^{51}$ erg into the ISM. Shocks, created by the explosion, disperse molecular clouds and compress the ambient medium. Supernova remnants are believed to be sites of particle acceleration and the origin of high-energy galactic cosmic rays, which mainly consist of hadrons and, in a few percent, leptons. If cosmic-ray protons accelerated by SNRs penetrate into high-density clouds, they give rise to strong $\pi^0$-decay gamma-ray emission. Therefore, SNRs that interact with molecular clouds are favourable objects for a detailed study of these gamma-ray spectra and can help to understand the production, acceleration, and propagation of cosmic rays. At the same time, SNR-cloud interactions drive slow shocks into these clouds at a velocity that relates to the shock velocity in the surrounding medium via the density contrast between cloud- and intercloud gas \cite{McKee75}. These shocks strongly cool through molecular emission and can be observed in the far-IR and sub-mm wavelength regime. The study and interpretation of this emission can yield valuable information on the supernova explosion and its impact on the surrounding medium as well as the molecular clouds themselves. This information is needed for the understanding of various astrophysical questions, such as the energy balance of the ISM in galaxies or triggered star formation. In particular, a thorough analysis of the dense matter distribution around SNRs is needed for a realistic modelling of gamma-ray emission in order to understand the origin and acceleration of cosmic rays. We observed four SNRs (W44, W28, 3C391, IC443) that are known to interact with ambient molecular clouds in molecular lines that are excited by the shock interaction. These observations are then interpreted using a shock model to gain understanding of the physical and chemical conditions within the interaction regions. In this article we will only summarise our results for one of these SNRs, namely W44, that are based on observations with the APEX telescope and were published in an extensive paper by \cite{Anderl14}. 

\section{Magneto-hydrodynamical shocks and numerical models}
Shocks are ubiquitous in the interstellar medium (ISM). They occur whenever large pressure gradients lead to fluid-dynamical disturbances that move at a velocity that exceeds the local sound speed. As shocks dissipate kinetic energy into heat, they give rise to strong cooling radiation that constitutes excellent diagnostics for the study of the conditions in the shocked gas. The interpretation of this radiation requires the application of detailed numerical shock models that comprises the description not only of the kinematics and multi-fluid hydrodynamics of gas and dust, but also gas-phase and dust-surface chemistry in a large regime of temperatures and densities, of dust physics, of heating- and cooling processes and of radiative transport. 

Different types of shocks can be distinguished depending, e.g., on the value of the local magnetosonic speed. Hydrodynamical shocks are faster than any signal speed in the shocked medium, so the preshock medium is not able to dynamically adapt to the shockwave before it arrives. Therefore, the fluid variables undergo a discontinuous jump in these so-called J-type shocks (see e.g. \cite{Hollenbach79}). On the other hand, for weaker shocks in a magnetized medium with a low degree of ionization, magnetosonic waves can travel ahead the disturbance. The ionized fluid is accelerated and heated by the magnetic precursor, whereas the neutral fluid is essentially unaffected by it. The resulting ion-neutral drift heats and accelerate the neutral fluid, while at the same time radiative emission cools the gas. Because heating and cooling take place simultaneously, the shock transition can become continuous. These shocks are hence called C-type (see e.g \cite{Draine80}). Finally, young shocks that have not reached a stationary state yet can be approximated by introducing a J-type discontinuity in a C-type flow at a point in the steady-state profile that is located increasingly downstream as the age of the shock advances, hence their designation as CJ-type \cite{Lesaffre04b}.

The model we used for the interpretation of the observed integrated spectral line intensities of CO (see Section \ref{Sec3}) consists of a one-dimensional, plane-parallel shock model combined with a radiative transfer module \cite{Gusdorf12}. The shock model solves the magneto-hydrodynamical equations in parallel with a large chemical network, including more than 100 species linked by over 1000 reactions, for stationary C- and J-type shocks (\cite{Flower03a}, \cite{Flower03b}) or approximated non-stationary CJ-type shocks.

\section{CO observations of W44}\label{Sec3}

\begin{figure} 
\includegraphics[width=\textwidth]{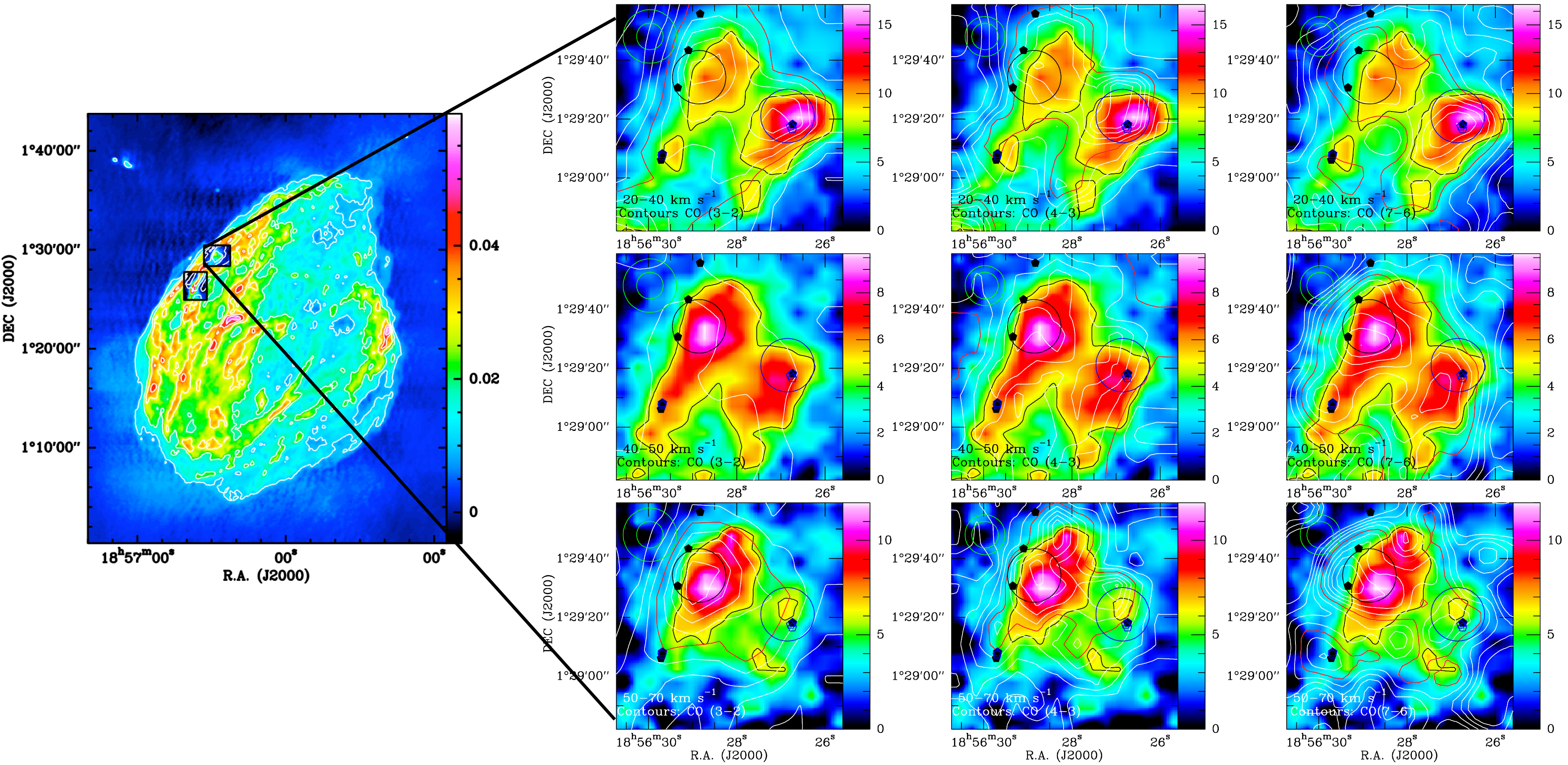} 
\caption{{\it Left:} Location of the fields W44E and F covered by our CO observations on the larger-scale radio continuum image at 1442.5 MHz, taken from \cite{Giacani97}. {\it Right:} Overlays of the velocity-integrated maps of CO (3--2) (left column), CO (4--3) (middle column), and CO (7--6) (right column) as white contours on the CO (6--5) emission (colour background) observed towards W44E with the APEX telescope. The maps are in their original resolution (9.0$''$ for CO (6--5), 13.5$''$ for CO (4--3), and 18.2$''$ for CO (3--2)), except for CO (7--6), which was smoothed to the resolution of CO (4--3) in order to improve the signal-to-noise. The intensity was integrated between 20--40 km s${}^{-1}$ (blue wing, top row), 40--50 km s${}^{-1}$ (ambient emission, middle row), and 50--70 km s${}^{-1}$ (red wing, bottom row). The wedge unit is K km s${}^{-1}$ in antenna temperature. The contours are in steps of 10\%. The half-maximum contours of the colour and contour maps are indicated in red and black, respectively. The blue and black circles indicate the position of the our shock modelling analysis. The APEX beam sizes of the observations displayed are given in the upper left corner of each map. The black and blue hexagons mark the positions of the OH masers observed by \cite{Claussen97} and \cite{Hoffman05}. At the assumed distance of 3 kpc the 9.0$''$ beam of CO (6--5) corresponds to a spatial resolution of 0.13 pc [Figure adapted from \cite{Anderl14}].
} 
\label{fig1}
\end{figure}

\begin{figure} 
\includegraphics[width=\textwidth]{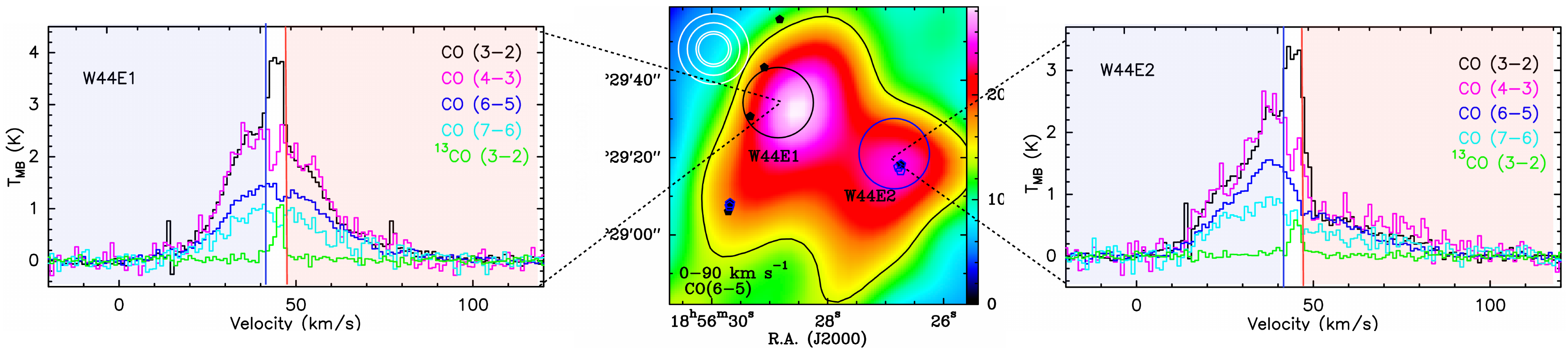} 
\caption{{\it Central Panel:} Positions of our analysis in W44E (W44E1: black circle, W44E2: blue circle) on the velocity-integrated CO (6--5) emission ($T_{\rm A}^*$ integrated between 0 km s${}^{-1}$ and 90 km s${}^{-1}$) convolved to the CO (3--2) angular resolution of 18.2$''$ (colour background) with half-maximum contour in black. The wedge unit is K km s${}^{-1}$. The blue and black circles indicate the APEX beam size of our observations in CO (3--2). The APEX beam sizes of our CO (3--2), CO (4--3), CO (6--5), and CO (7--6) observations are also provided (upper left corner). The black and light blue hexagons mark the position of the OH masers observed by \cite{Claussen97} and \cite{Hoffman05}. {\it Left Panel:} Spectra observed in the position W44E1 (in $T_{\rm MB}$), CO (3--2), black; CO (4--3), pink; CO (6--5), dark blue; CO (7--6), light blue; ${}^{13}$CO (3--2), green. The CO spectral resolution is 1.0 km s${}^{-1}$ for all lines. The background colours indicate the minimum blue and red integration intervals. {\it Right panel:} Spectra observed in the position W44E2, same colours as in the left panel [Figure adapted from \cite{Anderl14}].} 
\label{fig2}
\end{figure}

The supernova remnant W44 (also known as G34.7--0.4, or 3C 392) belongs to the class of so-called mixed-morphology supernova remnants. This term refers to SNRs with centrally concentrated X-ray emission and a shell-like radio morphology. W44's close distance of 3 kpc makes it a popular object for SNR studies. W44 is believed to be in a radiative phase over much of its surface given its estimated age of  $2 \times 10^4$~years and the observed cooling radiation. 
The flattened shell structure of the SNR in the northeast hints at an encounter with a dense molecular cloud. This interaction has been confirmed by several independent observations, e.g. emission from shock-tracing molecules and fine-structure line emission (e.g. \cite{Seta98}, \cite{Reach00}, \cite{Neufeld07}) as well as OH maser emission (\cite{Claussen97} and \cite{Hoffman05}). 

W44 is an important site for the study of cosmic-ray production, acceleration, and propagation based on the observations of $\gamma$-rays stemming from the interaction of cosmic-rays and the interstellar medium. These interactions include high-energy electron bremsstrahlung and inverse Compton scattering of the leptonic cosmic-ray component as well as proton-proton collisions of the hadronic component creating neutral pions, whose decay generates $\gamma$-rays. 
W44 is the first SNR where the spectral continuum below 200 MeV could be attributed to neutral pion emission, as established by the gamma-ray observatories AGILE and {\it Fermi} (\cite{Giuliani11}, \cite{Ackermann13}).

We observed two regions (W44E and W44F, see Fig. \ref{fig1}), in the northeastern flattened part of the radio shell in CO and $^{13}$CO with the Atacama Pathfinder Experiment (APEX telescope, \cite{Guesten06}). These regions show various signs of the interaction of the SNR and a molecular cloud. In particular, \cite{Claussen97} and \cite{Hoffman05} detected OH maser features at 1720 MHz. The right part of Fig. \ref{fig1} shows our maps of integrated CO emission in CO (6--5) with contours of integrated CO (3--2), (4--3), and (7--6) (respectively 1st, 2nd, and 3rd column) emission overlayed towards W44E, for the blue wing (top row, 20--40 km s${}^{-1}$), the red wing (bottom row, 50--70 km s${}^{-1}$), and the ambient velocity regime (central row, 40--50 km s${}^{-1}$). The positions of OH masers detected towards W44E are also displayed. In the following, we will only focus on our observations of W44E. The corresponding observations and results for W44F can be found in \cite{Anderl14}.
As shown in Fig. \ref{fig1}, the emission distribution in all CO maps peaks behind or near the line of masers, which trace the shock front moving eastward into the ambient molecular gas \cite{Claussen97}. There is another emission peak towards the interior of W44 close to the second masering region distinct from the delineated masers to the east. This peak is mostly seen in the blue-shifted velocity regime. Here the spatial displacement between maser and CO emission is less clear than for the other peak.
 
For our shock analysis, we aimed at choosing positions of clearly shocked gas, defined as positions of maximum CO emission in a given velocity range for a maximum number of observed transitions. Therefore, we first determined the local maxima of CO emission in all of our transition maps in the various velocity regimes. We then chose the positions of our analysis based on these maxima. In the centre panel of Fig. \ref{fig2}, the two positions resulting from this procedure are displayed: the one close to the line of masers tracing the outer shock front (black circle, denoted as W44E1) and the one towards the interior (blue circle, denoted as W44E2). The coordinates of W44E1 and W44E2 are (${\rm R.A.}_{[\rm J2000]}=$18$^{\rm h}$56$^{\rm m}$28.9$''$, ${\rm Dec}_{[\rm J2000]}=$01$^\circ$29$'$34.2$''$) and (${\rm R.A.}_{[\rm J2000]}=$18$^{\rm h}$56$^{\rm m}$26.9$''$, ${\rm Dec}_{[\rm J2000]}=$01$^\circ$29$'$21.0$''$), respectively. 
The individual spectra towards our positions of analysis in W44E, as shown in Fig. \ref{fig2}, exhibit a complex structure. In CO (3--2), two components are seen: there is a narrow component tracing the cold ambient gas and a broad component due to emission from shocked gas. In W44E1, the lines in CO (4--3), (6--5), and (7--6) show self-absorption features in the range between 43 km s${}^{-1}$ and 46 km s${}^{-1}$. The shock-broadened lines are asymmetric but similar in shape among all transitions. 

\section{Shock analysis}

All spectra we have extracted towards the positions of maximum CO emission in W44E show broad line wing emission around a central velocity consistent with the cold ambient gas (see Fig. \ref{fig2}). Probably these profiles arise from a superposition of different shock components propagating into the dense gas of molecular clumps, which are nearly impossible to disentangle. In order to still get significant constraints on the dominant shock features and environmental conditions, we separated the profiles into a blue and a red lobe and applied our shock analysis to each of these velocity domains independently. First, spectral line energy distributions (SLEDs, integrated line intensities for all transitions) were derived for both lobes separately. Filling factors, which were assumed to be the same for all transitions, were estimated based on the half-maximum contours in the maps of CO (6--5). The errors in the integrated intensities include the uncertainty of these filling factors as well as the uncertainties stemming from the integration intervals, the r.m.s. noise fluctuations of the profiles and the baseline subtraction. The observations were then compared with a grid of more than 1000 integrated intensity diagrams obtained for C- and CJ-type shock models using a $\chi^2$ routine. The fits were based on the most unambiguous shock-tracing lines CO (7--6) and CO (6--5), which are expected neither to be optically thick nor to be affected by self-absorption. 

\begin{figure}
\begin{center}
\includegraphics[width=0.45\textwidth]{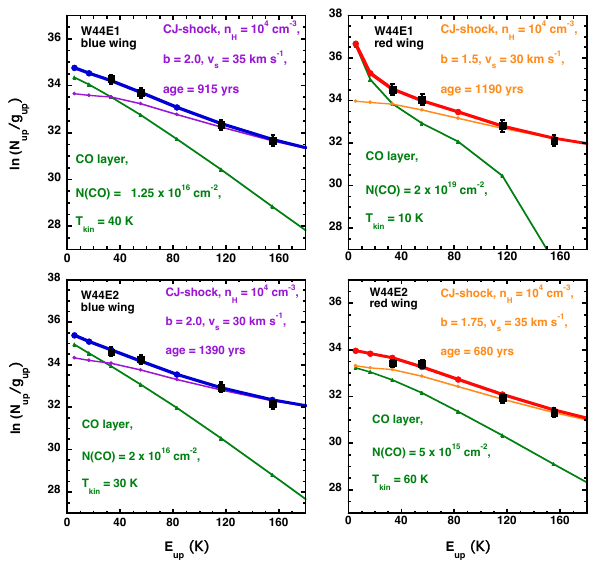} 
\caption{Best-model comparisons between CO observations and models for all positions in W44E (W44E1: top row, W44E2: bottom row) for both the blue and red shock components, respectively. Observations are marked by black squares, the best-shock models are displayed in violet line (blue lobe) and orange line (red lobe) and diamonds, the CO layer that we used to compensate the ambient emission affecting the CO (3--2) and CO (4--3) transitions in green line and triangels, and the sum of ambient and shock component in blue line (blue lobe) and red line (red lobe) with circles [Figure adapted from \cite{Anderl14}].     } 
\label{fig3}
\end{center}
\end{figure}
 
\section{Results}
The observed integrated intensities of CO in the unambiguously shock tracing transitions (7--6) and (6--5) towards W44E are only compatible with non-stationary CJ-shocks, as revealed by our $\chi ^2$ comparison with the grid of shock models. This result seems reasonable, given the presumably young age of the shocks in the molecular clumps close to the edge of the remnant (being much younger than the assumed SNR's age of 20,000 years). For all best-fitting models, the pre-shock density is found to be 10${}^{4}$ cm${}^{-3}$, consistent with previous estimates for the dense clump component (e.g. \cite{Reach00}). A general constraint on possible shock velocities is imposed by the total, blue- to redshifted velocity extent of the CO lines towards W44E, which is $\sim$60 km s${}^{-1}$. Obeying this constraint we obtain projected shock velocities of 35 and 30 km s${}^{-1}$ for the blue and the red lobe in W44E1, respectively, while in W44E2 we get 30 and 35 km s${}^{-1}$ for the blue and the red lobe. The pre-shock magnetic field component perpendicular to the shock layers in our models is constrained to $B =$ 200 $\mu$G for the blue lobes and 150 or 175 $\mu$G for the red lobes in W44E1 and W44E2. This difference could be interpreted as a projection effect, as a slight inhomogeneity of the medium or might also be due to uncertainties in the analysis. The modelled ages scatter around 1000 years.

As also found in our study of W28F \cite{Gusdorf12}, the emission in CO (3--2) and CO (4--3) is underestimated by all fitting shock models. Given the one-dimensional geometry of our models, this is not too surprising because the dense and cold postshock gas is not accurately modelled. Furthermore, the existence of OH masers hints at the presence of dense postshock gas that is due to a shock transverse to the line of sight and thus different from the line-wing shock emission we consider in our shock models. 
We therefore calculated a grid of CO layers with the LVG module in homogeneous slab mode. Based on a $\chi^2$ comparison with this grid, a layer of CO was then added to the shock models. The temperatures of the layers that were found to complement the shock fits are at temperatures between 10 K and 60 K with densities always higher than the obtained preshock density of 10${}^4$ cm${}^{-3}$. The CO column densities vary between 5$\times$10$^{15}$ cm${}^{-2}$ and 2$\times$10$^{19}$ cm${}^{-2}$. The quality of the fits for the positions in W44E can be seen in Fig. \ref{fig3} and is satisfying in all cases. The density and temperature values of the added gas layers are typical of postshock gas. In view of the geometric uncertainties, we however note that a more detailed interpretation of these layers (e.g. possible contamination by unshocked ambient / foreground gas) has to remain open.

These results could be used to estimate the energy and moment input for the interstellar medium. The shocked mass per beam amounts to $\sim$1 $M_{\odot}$ for the models in W44E. The total energy input in one beam, given as the sum of the blue- and the red-lobe shock contributions, amounts to $\sim$1$\times$10$^{46}$ erg. The values of the momentum input as sum of the red- and blue-shifted velocity lobes in one beam lies between $\sim$4$\times$10$^{6}$ $M_{\odot}$ cm s$^{-1}$ and $\sim$7$\times$10$^{6}$ $M_{\odot}$ cm s$^{-1}$. Furthermore, our best-fittig shock models allow to derive the integrated CO fluxes for all rotational transitions up to an upper rotational quantum number of 40. This information can be used to remove modelling degeneracies in our analysis by observations of higher-lying CO transitions and can be useful in the understanding of the shock contribution to the overall CO emission of nearby galaxies (cf. \cite{Kamenetzky14}). 

A final remark must be passed on the fact that our method relies on a very simplified geometry of the shocked region. We split the whole spectral line into blue- and red-shifted parts without completely omitting the central line emission. The real situation, however, is most certainly much more complex. The broad line wings might  stem from the projected wings of a bow-shock, with the ambient medium being compressed and pushed aside. Our analysis can only be applied to shock components propagating parallel to the line of sight and therefore gives averaged information on the shock conditions only in the gas being pushed towards or away from us. However, the derived properties, as long as they are non-directional such as the total magnetic field strength or the density, should still be valid despite our simplifying assumptions. We also note, that the high level of microphysical detailedness included in our shock model cannot be maintained within fully two- or three-dimensional models, while the use of pseudo-multidimensional models (\cite{Kristensen08}, \cite{Gustafsson10}), synthesized from one-dimensional shocks, would add a set of additional, barely constrained parameters.

\section{Summary}
\begin{itemize}
\item Two regions in the supernova remnant W44 were mapped with the APEX telescope in $^{12}$CO (3--2), (4--3), (6--5), (7--6) and $^{13}$CO (3--2). The spectra in the positions of maximum emission show narrow features of ambient emission and a broad component of shock emission.
\item A comparison of the integrated intensities of CO (7--6) and (6--5) in these positions with a large grid of shock models combined with a radiative transfer module yielded constraints on the shock characteristics and pre-shock conditions in the observed regions. Furthermore, the SNR's impact on its environment in terms of mass, momentum and energy dissipation was quantified. 
\item For the first time, young non-stationary shock models were shown to be consistent with the shock emission in W44E and W44F. 
\item Based on the best-fitting shock models, the pre-shock density was estimated as 10$^4$ cm$^{-3}$, the magnetic field strength was found to be $\sim$200 $\mu$G, and the shock velocities of the best-fitting models are found to lie between 20 and 35 km s$^{-1}$. 
\end{itemize}

\end{document}